\documentclass[fleqn,usenatbib]{mnras}

\usepackage{newtxtext,newtxmath}

\usepackage[T1]{fontenc}

\DeclareRobustCommand{\VAN}[3]{#2}
\let\VANthebibliography\thebibliography
\def\thebibliography{\DeclareRobustCommand{\VAN}[3]{##3}\VANthebibliography}

\usepackage{graphicx}	
\usepackage{amsmath}	

\usepackage{caption}
\usepackage{subcaption}

\usepackage{gensymb}

\title[Precessing jets from a neutron star]{Relativistic precessing jets powered by an accreting neutron star}

\author[F. J. Cowie et al.]{F. J. Cowie,$^{1}$\thanks{E-mail: fraser.cowie@physics.ox.ac.uk}
R. P. Fender,$^{1,2}$
I. Heywood,$^{1,3}$
A. K. Hughes,$^{1}$
K. Savard,$^{1}$
P. A. Woudt,$^{2}$
\newauthor F. Carotenuto,$^{4}$
A. J. Cooper,$^{1}$
J. van den Eijnden,$^{5}$
K. V. S. Gasealahwe,$^{2,6}$
S. E. Motta,$^{1,7}$
P. Saikia$^{8}$
\\
$^{1}$Department of Physics, University of Oxford, Denys Wilkinson Building, Keble Road, Oxford OX1 3RH, UK \\
$^{2}$Department of Astronomy, University of Cape Town, Private Bag X3, 7701 Rondebosch, South Africa\\
$^{3}$Centre for Radio Astronomy Techniques and Technologies, Department of Physics and Electronics, Rhodes University, PO Box 94,
Makhanda, 6140, South Africa\\
$^{4}$INAF, Osservatorio Astronomico di Roma, Via Frascati 33, I-00078 Monte Porzio Catone, Italy\\
$^{5}$Anton Pannekoek Institute for Astronomy, Universiteit van Amsterdam, Science Park 904, 1098, XH, Amsterdam, The Netherlands\\
$^{6}$South African Astronomical Observatory, P.O. Box 9, 7935
Observatory, South Africa \\
$^{7}$Istituto Nazionale di Astrofisica, Osservatorio Astronomico di Brera, via E. Bianchi 46, 23807 Merate (LC), Italy \\
$^{8}$Center for Astro, Particle and Planetary Physics, New York University Abu Dhabi, PO Box 129188 Abu Dhabi, UAE \\
}

\date{Accepted XXX. Received YYY; in original form ZZZ}

\pubyear{2025}

\begin{document}
\label{firstpage}
\pagerange{\pageref{firstpage}--\pageref{lastpage}}
\maketitle

\begin{abstract}

\noindent Precessing relativistic jets launched by compact objects are rarely directly measured, and present an invaluable opportunity to better understand many features of astrophysical jets. In this Letter we present MeerKAT radio observations of the neutron star X-ray binary system (NSXB) Circinus X-1 (Cir X-1). We observe a curved S-shaped morphology on $\sim 20''$ $(\sim1\:\text{pc})$ scales in the radio emission around Cir X-1. We identify flux density and position changes in the S-shaped emission on year timescales, robustly showing its association with relativistic jets. The jets of Cir X-1 are still propagating with mildly relativistic velocities $\sim1\:\text{pc}$ from the core, the first time such large scale jets have been seen from a NSXB. The position angle of the jet axis is observed to vary on year timescales, over an extreme range of at least $110\degree$. The morphology and position angle changes of the jet are best explained by a smoothly changing launch direction, verifying suggestions from previous literature, and indicating that precession of the jets is occurring. Steady precession of the jet is one interpretation of the data, and if occurring, we constrain the precession period and half-opening angle to $>10$ years and $>33\degree$ respectively, indicating precession in a different parameter space to similar known objects such as SS~433.

\end{abstract}





\section{Introduction}


Relativistic jets are amongst the most important phenomena in astrophysics, associated with accretion on a vast range of scales from gamma-ray bursts to supermassive black holes. Observations of different astrophysical jets have shown that in some cases the direction in which they are launched can change with time. On large spatial scales this phenomenon is used to explain the morphology of certain radio galaxies \citep{horton_2020}, but it is only on smaller scales where the change in the jet launch axis with time can be directly observed. This has only been seen in a handful of systems, but over a range of size scales, from the supermassive black hole in M87 \citep{cui_2023} to the stellar mass compact object in the microquasar SS~433 (\citealt{fabian_1979}, \citealt{milgrom_1979}, \citealt{abell_1979}).



Astrophysical jet systems with measurable changes in jet launch direction, though rarely directly observed, are invaluable for probing key unknowns in jet physics. They allow conditions close to the jet launching region to be understood, and allow for insight into the mechanism powering the jet. Furthermore, jets which change direction are more efficient at transferring energy to the ambient medium due to the increased interaction area \citep{monceau_2014}. If jets which change direction are a widespread phenomenon this then has consequences for particle acceleration (\citealt{britzen_2019}, \citealt{hess_2024}) and feedback (\citealt{vernaleo_2006}, \citealt{li_2014}). Steady precession of the jet launch axis is the common model invoked to explain changing jet directions, whether due to Lense-Thirring precession of the inner accretion disc (\citealt{lense_1918}, \citealt{liska_2018}), or tidal effects of a binary companion on a misaligned accretion disc \citep{larwood_1998}. Jets can show changing directions from other effects, such as motion of the compact object through the ISM/IGM \citep{heinz_2008}, or from interactions with a strong wind \citep{yoon_2015}, but in these cases the intrinsic jet \textit{launch} direction is not changing. 

X-ray binaries (XRB) are a unique laboratory for probing accretion and relativistic jets, due to their stellar scale size leading to evolution on human accessible timescales when compared to their supermassive counterparts. The canonical example of a precessing jet is that of the XRB SS~433 (\citealt{margon_1984} and references therein), precisely measured using radio imaging (\citealt{hjellming_1986}, \citealt{blundell_2004}) and spectral lines (\citealt{margon_1984}, \citealt{eikenberry_2001}). Position angle variations in the jet from the black hole XRB V404 Cyg have also been observed and attributed to Lense-Thirring precession \citep{miller-jones_2019}. Hints of a narrow precession in the jets of 1E 1740.7-2942 \citep{escamilla_2015} and Sco X-1 \citep{fomalont_2001b} have also been reported. In many cases the observation of certain quasi-periodic oscillations (QPOs) in XRB systems has been used to suggest the presence of a precessing jet (\citealt{ingram_2019}, \citealt{zhang_2023}, \citealt{yang_2024}). Identifying more of these rarely observed precessing jets, with high quality observations, is vital to better understand this phenomenon and will allow these systems to be better used as accretion and jet laboratories.


Circinus X-1 (Cir X-1) is a peculiar neutron star X-ray (NSXB) binary system which defies conventional classification. Discovered in 1969 \citep{margon_1971}, it has been monitored and studied for over 50 years in X-ray, optical, infrared, and at radio wavelengths (e.g. \citealt{yu_2024}, \citealt{johnston_2001}, \citealt{glass_1994}, \citealt{whelan_1977}). At all these wavelengths it has a periodic lightcurve of approximately $16.5$ days (\citealt{tominaga_2023}, \citealt{johnston_2016}, \citealt{glass_1978}, \citealt{calvelo_2012a}). The X-ray properties of Cir X-1 are similar to low mass NSXBs, with Z and Atoll behaviour \citep{soleri_2009}, and QPOs \citep{boutloukos_2006}. Type \MakeUppercase{\romannumeral 1} X-ray bursts from the system confirmed a neutron star accretor (\citealt{tennant_1986}, \citealt{linares_2010}). Cir X-1 is surrounded by its natal supernova remnant (SNR), known as the Africa nebula \citep{gasealahwe_2025}, indicating the system is less than 4600 years old, making it the youngest known XRB \citep{heinz_2013}. The nature of the companion star is uncertain (\citealt{jonker_2007}, \citealt{johnston_2016}, \citealt{schulz_2020}). Observations of dust scattering rings suggest a distance of $\sim9.4\:\text{kpc}$, implying Cir X-1 frequently exceeds the Eddington limit for a $1.4 \: \textup{M}_\odot$ neutron star. 


Overall, the system can be modelled as a neutron star in an eccentric orbit with a companion star, where at periastron increased mass transfer and accretion rate leads to an X-ray flare and the launch of a transient relativistic jet (\citealt{murdin_1980}, \citealt{moin_2011}). The launch of the jet leads to a radio flare which shows an optically thick to thin synchrotron evolution, consistent with an expanding synchrotron emitting plasma (\citealt{haynes_1978}, \citealt{haynes_1979}, \citealt{calvelo_2012a}). There is also some evidence of radio flaring at apastron (\citealt{fender_1997}, \citealt{tudose_2008}). VLBI observations of Cir X-1 around periastron showed a resolved symmetric jet-like structure at milliarcsecond (mas) scales, constraining the jet collimation angle to be $<20\degree$ \citep{miller-jones_2011b}. These observations suggested a mildly relativistic jet inclined close to the plane of the sky. However, other observations suggest the presence of an unseen ultra-relativistic flow (URF), launched close to the plane of the sky during flaring of the source, energising downstream material in the jet \citep{fender_2004}, behaviour also seen in the XRB Sco X-1 \citep{fomalont_2001}. However, the nature of the recurrent flaring of Cir X-1 could potentially complicate the interpretation of these core-lobe time delays used to infer the presence of this URF.

Observations in the literature show evolution in the morphology and luminosity of the jets over timescales from days to years, including tentative variation in the jet position angle direction \citep{tudose_2008}. \cite{calvelo_2012b} presented the first millimetre wavelength observations of the source, finding morphology and flux density variations which again suggested that the jet position angle had changed from previous observations. \cite{coriat_2019} showed that the jet position angle appears to change with distance from the core. Long term secular evolution has also been observed in the radio and X-ray luminosities (\citealt{armstrong_2013}, \citealt{yu_2024}). Potential jet interaction regions with the ambient medium have been found aligned with the historical jet axis. These include wide angled "cap" structures within the bounds of the SNR \citep{sell_2010} and asymmetric bubbles where the jet has punched through the SNR \citep{gasealahwe_2025}.

Currently it is  difficult to imagine a single simple model which can explain the various and often contradicting phenomenology of Cir X-1 and its jets. The many peculiar features of the jets of Cir X-1, strongly motivates their further detailed study.


\section{Observations \& data processing}\label{sec:obs}


We observed Cir X-1 and the surrounding field using both the L-band (856-1712~MHz) and the S-band (1.75-3.5~GHz, 875~MHz instantaneous bandwidth) receivers on the MeerKAT radio telescope (\citealt{meerkat}, \citealt{s_band}). These observations were made as part of the ThunderKAT \citep{thunderkat}, and X-KAT programmes (MeerKAT Proposal ID: SCI-20230907-RF-01). 

In time order the observations taken were: 45 minute L-band (1.3 GHz) observation on 27/10/2018; 3x15 minute S2-band observations (2.6 GHz) from the 11/08/2023 with daily cadence; 15 minute L-band observation on 04/02/2024; 60 minute S4-band (3.1 GHz) observation on 26/10/2024; 60 minute S2-band observation on 31/05/2025. The length of observation refers to the total on source time, composed of scans up to 30 minutes long sandwiched by 2 min secondary calibrator (J1427–4206) scans. Each observation began and ended with a 5-minute scan of J1939–6342, the primary calibrator.

The observational data were reduced using the semi-automated pipeline \textsc{oxkat}\footnote{\url{https://github.com/IanHeywood/oxkat}} \citep{oxkat}. The software packages used in this pipeline were accessed using \textsc{singularity} for software containerisation \citep{singularity}. First generation calibration (applying corrections from the primary and secondary calibrators, see \citealt{heywood_2022}) and iterative RFI flagging was done using \textsc{casa} \citep{casa}. The target field was flagged further using \textsc{tricolour} \citep{tricolour} and then imaged using \textsc{wsclean} \citep{wsclean}. A deconvolution mask was manually made and iteratively improved through further imaging and use of the mask making tool \textsc{breizorro} \citep{breizorro}, before phase and delay self calibration were performed using \textsc{cubical} \citep{cubical}. 


Joined frequency channel deconvolution using 8 channels and a Briggs weighting of -0.3 was used \citep{briggs}. For the images at S2 and S4-band an inner Tukey taper from $0$ to $10^4\lambda$ was applied to the weighting of the visibilities in the uv-plane and at L-band an inner Tukey taper from $0$ to $5\times10^3\lambda$ was used. Tapering allowed for down-weighting of the diffuse background emission from the SNR. In all images a known background source to the SE was subtracted (see \citealt{sell_2010}, \citealt{calvelo_2012a}, \citealt{coriat_2019}). The source subtraction was done by fitting an elliptical Gaussian with fixed size given by the restoring beam, plus a flat background, allowing the flux density of the Gaussian component and background to vary, and allowing the position of the component to vary only $\pm0.5$ pixels from the known position of the background source from our high resolution uniformly weighted images. We investigated systematic astrometric errors by examining the variability in the best-fit positions of 10 background sources of varying flux density relative to their mean positions. Combining the offsets across the sample, we find a standard deviation of $\sim0.07''$, which is significantly smaller than our adopted systematic uncertainties throughout this letter. This indicates that our interpretations are robust under the conservative assumption that astrometry is limited to 10\% of the synthesised beam (i.e., $\sim0.3''$). The source subtraction was implemented using the {\sc{astropy}} \citep{astropy} and {\sc{lmfit}} packages \citep{lmfit}. In some images a point source was subtracted using the same process from the known core position of Cir X-1 \citep{moneti_1992}, in order to more clearly show any resolved underlying structure. We note that the position of the core of Cir X-1 in our high resolution images agrees with the known position of  the optical counterpart \citep{moneti_1992}. Throughout this Letter a distance of $9.4\:\text{kpc}$ is assumed, and the uncertainty on this distance \citep{heinz_2015} does not affect any conclusions.

\section{Results}\label{sec:results}

\begin{figure}
 \includegraphics[width=\columnwidth]{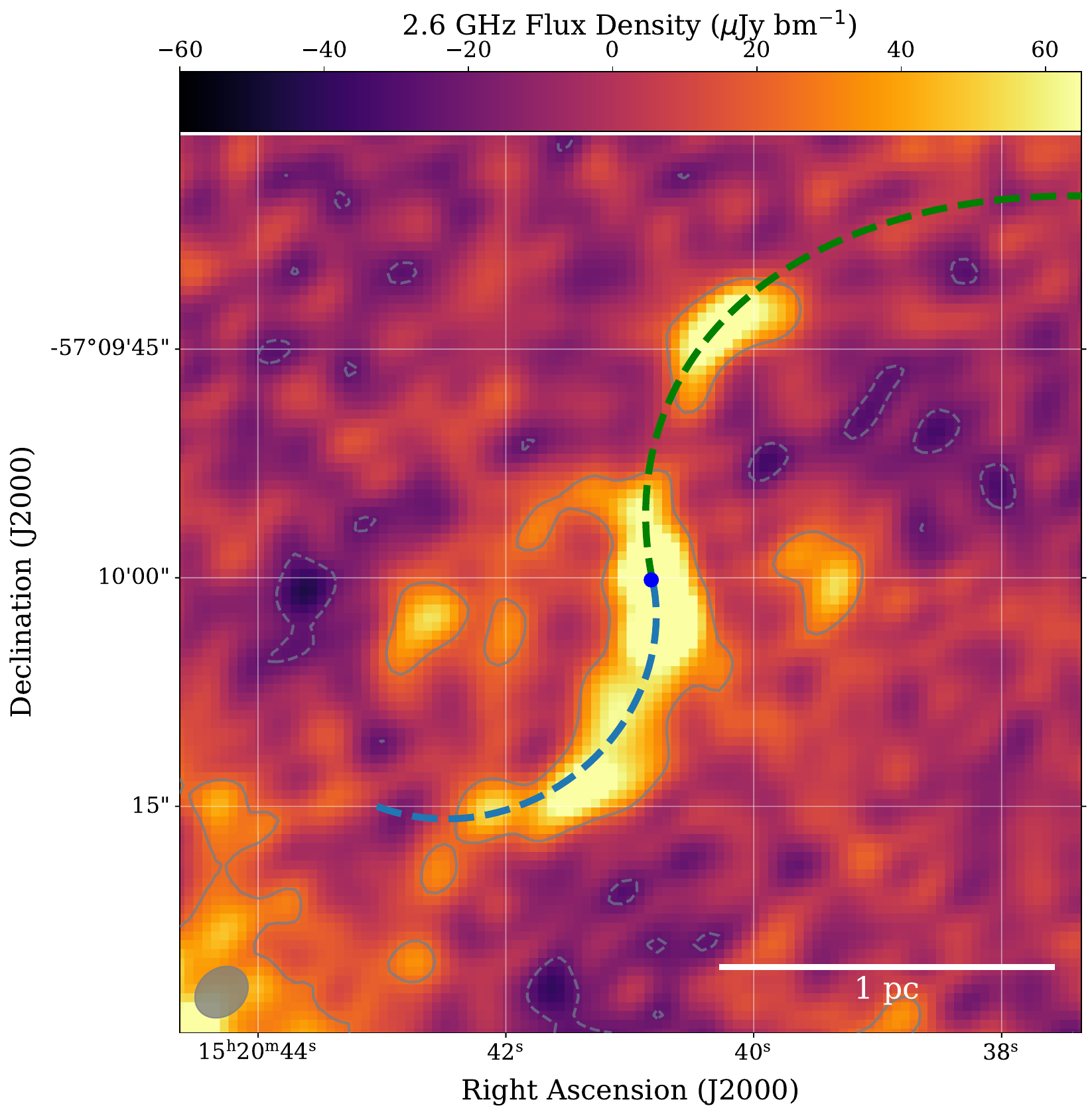}
 \caption{S2-band image of the jets of Cir X-1 in August 2023, the known background source to the SE has been subtracted. The core of Cir X-1 is marked by a blue dot. This image is the result of combined imaging of 3x15 minute observations taken within a 48 hour period in August 2023. The dashed line is a fit to a ballistic precession model. The noise level in the region surrounding Cir X-1 is $8\:\mu\text{Jy beam}^{-1}$, $-3\sigma,3\sigma$ contours are shown.}
 \label{fig:core_no_sub}
\end{figure}

\begin{figure}
 \includegraphics[width=\columnwidth]{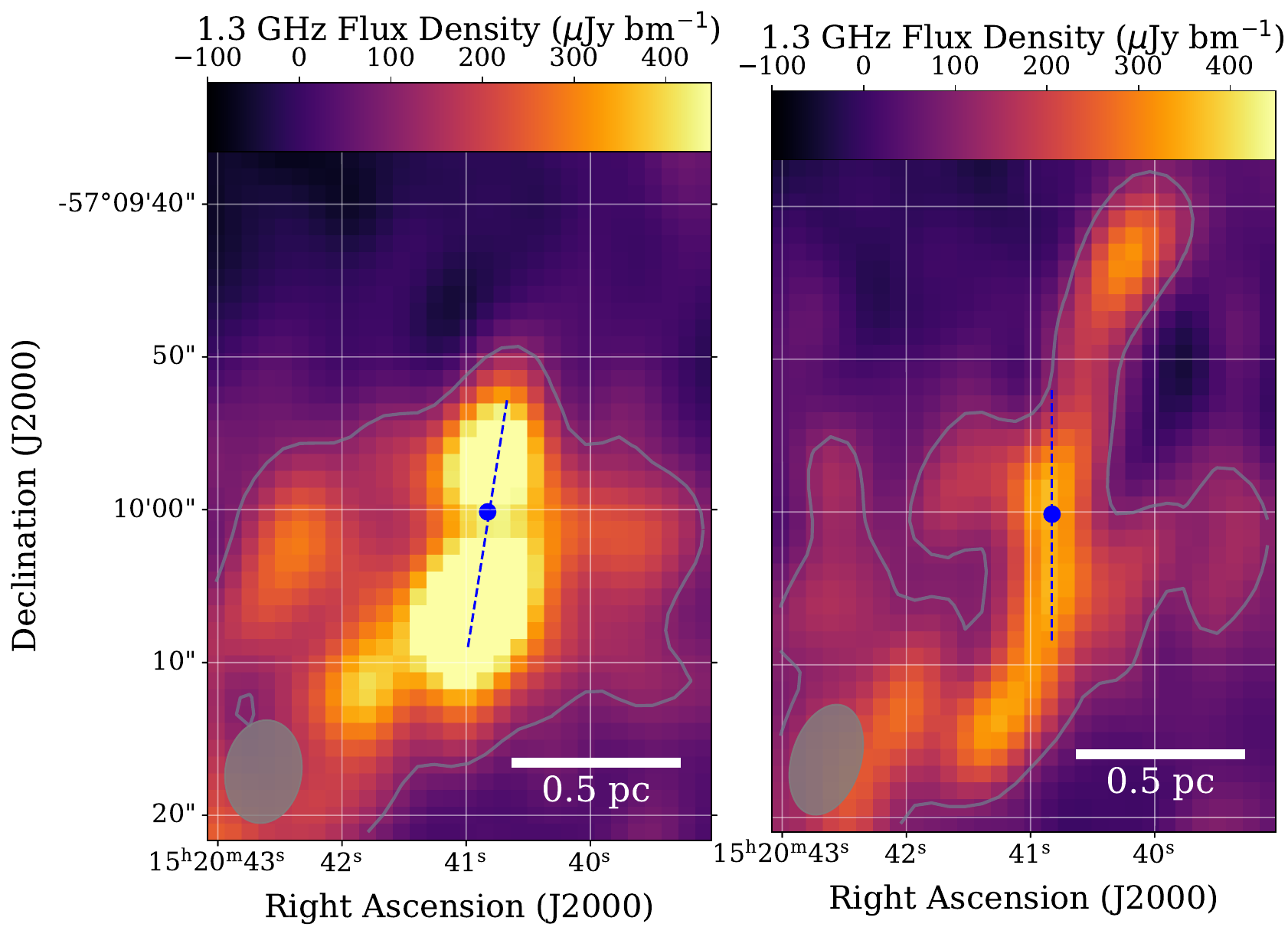}
 \caption{Left: L-band image of the jets of Cir X-1 in 2018 the core (marked by a blue dot) and SE background source have been subtracted. Right:  L-band image of the jets of Cir X-1 in 2024 with the same subtractions. The position angle of the core emission, $-9\pm3\degree$ E of N in 2018 and $\sim0\degree$ E of N in 2024, is indicated by a dotted line. Between the two images, there is variability in the position angle, morphology and flux density of the emission close to the core and at larger distances. The noise level in the region surrounding Cir X-1 is $35\:\mu\text{Jy beam}^{-1}$ in both images, $-3\sigma,3\sigma$ contours are shown.}
 \label{fig:L_band_comp}
\end{figure}

\begin{figure}
 \includegraphics[width=\columnwidth]{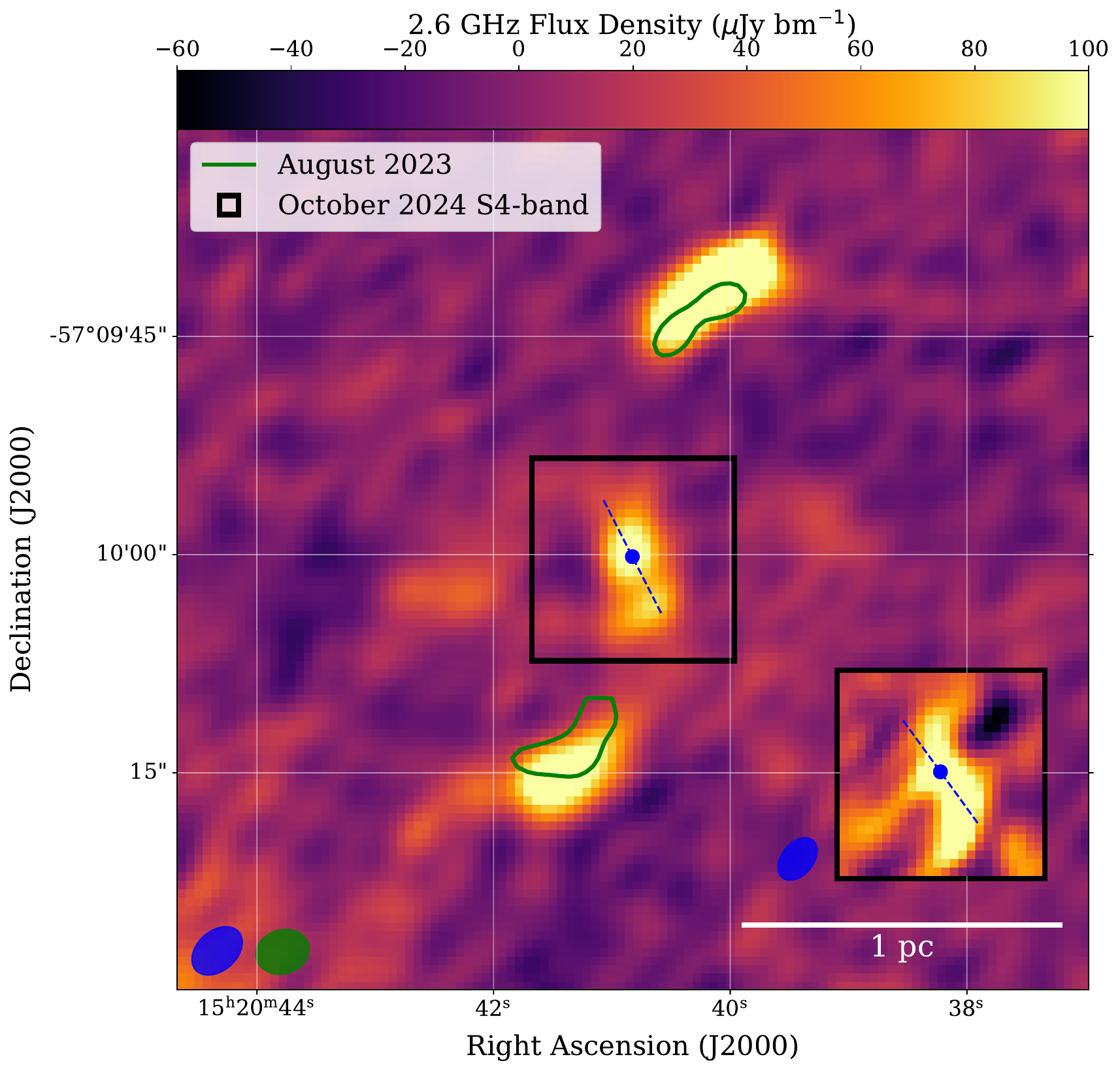}
 \caption{S2-band image of the jets of Cir X-1 in May 2025, with known background source and core subtracted. The lower right inset shows the higher resolution S4-band view of the jets close to the subtracted core in October 2024, where a miniature S-shape can be seen. In green are $10\sigma$ contours from the observation shown in Figure~\ref{fig:core_no_sub}, illustrating movement of components in the S-shape. The core is marked by a blue dot and the position angle of emission surrounding the core is $36\pm4\degree$, indicated by a dotted line. Contours from the central part of the image have been removed for clarity. The noise level in the region surrounding Cir X-1 is $9\:\mu\text{Jy beam}^{-1}$ in both the image and the inset. The blue ellipse in the bottom left represents the beam in the 2025 image displayed, while the green ellipse represents the beam for the 2023 observation, for which contours are shown. The blue ellipse next to the inset region is the beam for the 2024 observation shown in the inset.}
 \label{fig:1yrlater}
\end{figure}

Figure~\ref{fig:core_no_sub} clearly shows that the radio emission surrounding Cir X-1 has a curved symmetric S-shaped morphology on parsec scales. This morphology is particularly clear in these 3x15 minute epochs taken in August 2023 as the orbital phase was $\sim0.8$ (using an updated radio ephemeris, Nicolson priv. comm.), meaning there is no/minimal radio emission from the core \citep{calvelo_2012a}. The lack of core flux density variability allows them to be combined during imaging despite being taken 48 hours apart. Other observations during the campaign, including longer contiguous observations, showed the same curved morphology but an increased core flux density meant that source subtraction of the core was required to see these features more clearly. Furthermore, we note that the radio emission within $\sim3''$ of the core of Cir X-1 is extended along an axis. The position angle of this core emission in Figure~\ref{fig:core_no_sub} is $13\pm3\degree$ E of N. The jet emission has had position angle measurements made over the past 20 years, but has never been observed at this angle before.

If the curved symmetric S-shape radio morphology is associated with the jets from Cir X-1 then we find some form of precession the most plausible scenario (see Section~\ref{sec:discussion}). Ballistic precession is one such geometric precession model, where the precession parameters are stable and no deceleration of the jet occurs. Figure~\ref{fig:core_no_sub} shows a fit to a ballistic precession model \citep{hjellming_1986}. Data points were placed evenly along the S-shape, and the differential evolution algorithm \citep{storn_1997}, implemented in {\sc{scipy.optimize}} \citep{scipy}, was used to minimise the 2D on-sky distance between data and model. The jet speed and geometric parameters describing the precession cone are detailed in Table~\ref{tab:preccession_params}. However, due to degeneracy these are not to be taken as the precession parameters (see Section~\ref{sec:discussion}).

It is clear that there is much complex radio structure close to Cir X-1, not all of which may be attributable to relativistic jets. However, we can show using additional data that the S-shaped radio morphology is unambiguously associated with relativistic jets. The left of Figure~\ref{fig:L_band_comp} shows the 2018 L-band observation with the core and background source subtracted. The right of Figure~\ref{fig:L_band_comp} shows the 2024 L-band observation with the same subtractions. Both the 2018 and 2024 data show a clear extended structure of emission close to the subtracted core. In the 2018 case the structure is bipolar. In 2024 on larger scales hints of the S-shape seen more clearly in Figure~\ref{fig:core_no_sub} are present. In 2018 the position angle of the core emission is $-9\pm3\degree$. This is significantly different from the position angle on a similar size scale of core emission measured in 2023, shown in Figure~\ref{fig:core_no_sub}. The position angle of the core structure in the 2024 L-band data is difficult to measure accurately due to low flux density when compared to background emission, but appears to be close to $0\degree$ E of N.

An important conclusion from the comparison of images at the same frequency in 2018 and 2024 in Figure~\ref{fig:L_band_comp} is the clear variability of flux density and structure of the emission both close to the core and on parsec scales. Notably, the northern and southern components of the S-shape are not present in 2018. While the cause of the variability close to the core may be short or long term, as these observations were taken at different phases in the binary orbit, the variability far from the core confirms that this emission is associated with Cir X-1, not the surrounding SNR.

We can then compare the S2-band observations in 2023 in Figure~\ref{fig:core_no_sub} to S2-band observations taken nearly 2 years later in 2025. Furthermore, we can also use the highest frequency and resolution S4-band observations from 2024 to investigate the structure of the emission very close to the core, after core subtraction. Figure~\ref{fig:1yrlater} shows the image from the S2-band 2025 observations, with an inset showing the core in the 2024 S4-band observations. Finally, $10\sigma$ contours from the 2023 observation shown in Figure~\ref{fig:core_no_sub} are plotted in order to illustrate movement of the northern and southern components of the S-shape between 2023 and 2025.

In order to measure the positional difference in the north (N) and south (S) components of the S-shape between 2023 and 2025 we use 2 methods as no simple image plane model is appropriate to fit. We first define two axes along which positional differences will be measured, one  for the N component and the other for the S component. This is done by connecting the brightest pixel in Figure~\ref{fig:core_no_sub} for each component and the core position. One measurement method is then to simply identify the brightest pixel in Figure~\ref{fig:1yrlater} in each component along the respective axis, and calculate the distance from the brightest pixel in Figure~\ref{fig:core_no_sub}. A more advanced method to measure the positional differences uses the gradient based method of Canny edge detection \citep{canny2} to identify the leading edge of the emission in 2023 and 2025. Then along the the respective axes the positional difference of the edge in 2023 and 2025 was measured. The edge detection was implemented using the {\sc{scikit-image}} package \citep{scikit_image}, and the parameters used to run the edge detection were a low threshold of $40~\micro\text{Jy bm}^{-1}$ and a high threshold of $70~\micro\text{Jy bm}^{-1}$. No Gaussian blur was applied to the image beforehand.

The brightest pixel method gave positional differences of  $2.1\pm0.4''$ and $1.8\pm0.4''$ for the N and S components respectively, where a positive difference in position corresponds to increasing distance from the core. The edge detection method gave a difference in position of $2.4\pm0.4''$ for both the N and S components. In all cases the uncertainties have been estimated by calculating the uncertainty which would be present on an elliptical Gaussian fit to a point source with the same SNR as the N and S components \citep{condon_1997}, including a 10\% systematic error (which dominates) to ensure the uncertainty estimates are conservative. Other studies using MeerKAT have found this systematic error to be significantly less than 10\% \citep{hughes_2024}, suggesting that the uncertainties on the position differences are likely very conservative. 

Our analysis shows strong evidence that the outer parts of the S-shaped emission are propagating away from the core of Cir X-1 with proper motions of approximately $3.5\pm0.6\:\text{mas day}^{-1}$ (taking the positional difference derived from the edge detection method). At a distance of $\sim9.4\:\text{kpc}$ \citep{heinz_2015}, this corresponds to an mildly relativistic apparent velocity of $0.19\pm0.04c$. Then making the usual assumption (which appears to hold for SS 433 and other sources e.g. \citealt{hjellming_1986}) that the brightest point in the jet is conserved, we can say this value is the apparent bulk velocity of the material within the jet. This result, combined with the observed variability, shows unambiguous evidence that the S-shaped emission is associated with a relativistic jet from Cir X-1, which is still propagating mildly relativistically out to parsec scales. 

We also observe strong variability in the flux density of the northern and southern components between the 2023 and 2025 observations at S2-band, with the northern jet component brightening by a factor of $>3$. While the large scale S-shape is overall fairly symmetrical, we see that in Figures~\ref{fig:core_no_sub} and \ref{fig:1yrlater}, for a given position angle, the N component of the S-shape is further from the core than the S component, by $\sim3''$. In Figure~\ref{fig:1yrlater}, following the S-shape morphology towards the south-east, we observe a tentative hint of an inflexion in the morphology. Future observations will be needed to confirm if this emission associated with the jets.


In Figure~\ref{fig:1yrlater} the position angle of the core emission in 2025 is $28\pm4\degree$. In the inset of Figure~\ref{fig:1yrlater}, showing higher resolution 2024 data, emission surrounding the core is not symmetric but shows an asymmetric miniature S-shaped morphology. Therefore, we measure the position angle of the core emission in this case by using the emission to the south-west of the core, which is brighter and more extended than the emission to the north. The measured position angle of the core emission in this case is $36\pm4\degree$. The significance of these residuals ($>80 \:\mu\text{Jy beam}^{-1}$) was examined by performing a similar core subtraction on background point sources of comparable flux density. Despite the test sources being off-axis (and thus more prone to calibration systematics), our tests revealed no comparable residual structure, strongly suggesting that the mini S-shaped morphology is intrinsic to the source and not an artefact. Measurements of the core emission position angle on arcsecond scales are not necessarily indicative of the jet launch axis position angle at that time, but instead measure the jet launch position angle at some time in the past. However, we can, by combining our position angle observations with an estimate of the jet velocity, and the large scale S-shape, build up an overall picture of the jets in the system. 

Since $2015$ the position angle of the jet launch axis on the sky has rotated from $\sim-10\degree$ to at least $\sim36\degree$ E of N. On average the position angle of the jet launch axis has changed by $\sim5\degree\:\text{year}^{-1}$. Archival observations of the position angle of core emission show variation between $\sim-75\degree$ and $\sim-35\degree$ E of N at 8.6~GHz on scales of $\sim2''$ \citep{tudose_2008}, a position angle of $\sim-68\degree$ E of N on $\sim20\:\text{mas}$ scales \citep{miller-jones_2011b}, and angles between $\sim-15\degree$ and $\sim-60\degree$, on scales of $\sim2''$ and $\sim10''$ respectively \citep{coriat_2019}. Therefore, we can concretely say that the projected jet launch axis on the sky, has varied by \textit{at least} $\sim110\degree$ over the past 25 years a rather extreme and unusual position angle change for a jet launch axis compared to other sources.



Finally, in all our observations we see faint diffuse emission (at a significance of $6\sigma$ in Figure~\ref{fig:core_no_sub}), oriented in an E-W direction, roughly perpendicular to the current jet launch axis, at a distance of $\sim10''$ ($\sim0.5\:\text{pc}$) and approximately symmetrical about the core. Tentative hints of variability, most obvious in Figure~\ref{fig:L_band_comp}, suggest that this emission may also be related to the relativistic jets.

\section{Discussion}\label{sec:discussion}


In Section~\ref{sec:results} we showed that there is a symmetric parsec scale S-shaped morphology in the relativistic jets of Cir X-1. There are several scenarios which can explain a such a jet: 

\begin{enumerate}
    \item The jet itself is straight but has fluid elements travelling along curved trajectories due to instabilities \citep{nikonov_2023}.

     \item The jet follows a curved path due to interactions with the surrounding medium \citep{miller_2009}.

    \item The jet travels in a straight line but the launch axis has changed over time.
\end{enumerate}

Scenario (i) is expected to produce sharp small bends in the jet (\citealt{perucho_2012}, \citealt{misra_2025}), this is disfavoured due to the absence of these effects, and the observation of a well collimated curved jet out to large scales relative to the size of the system. The ambient environment of Cir X-1 is likely to be complex but low density in comparison to the ISM given the SNR. However, (ii) is disfavoured due to the requirement of the medium to be both inhomogeneous and symmetric being unlikely and it being unable to explain the time evolution of the core emission axis. This leaves scenario (iii) as the most likely. A caveat is that if a sufficiently powerful wind exists, with a changing launch direction, then jets could be redirected by this wind, creating the appearance of a changing jet launch direction. This mechanism has been invoked to explain the precession of the jet in SS~433 \citep{begelman_1980}, but would require Cir X-1 to have a remarkably powerful wind-like outflow, and therefore have a hyper-Eddington mass transfer rate.



We were able to reproduce the observed S-shape morphology using a simple ballistic jet model (no deceleration; \citealt{hjellming_1986}) in which the jet launch direction precesses over time. The model provides a good qualitative match to the morphology, but the similarity between a fast jet precessing on a short timescale and a slower jet on a longer timescale, the degeneracy between the opening angle and the position angle of the precession cone, and the coupling between intrinsic jet velocity and the line-of-sight inclination of the precession cone render the fitted parameters largely unconstraining. These degeneracies arise due to our observational data being limited to a small window of time relative to the precession period, and the lack of identification of an inflexion point in the S-shape, which would identify the edge of the precession cone.

Furthermore, in the case of Cir X-1 we note several features that may make this simple ballistic precession model unsuitable. To begin with, the measured combined proper motion of the parsec scale jets is significantly less than the value measured from VLBI observations \citep{miller-jones_2011b}. This could be either due to different inclination angles relative to the line of sight, or a reduction in the jet speed, likely due to deceleration on parsec scales. We know that the S-shaped emission on large scales must be caused by in situ particle acceleration as the initial synchrotron flare fades on day timescales before the jet propagates to large distances. This particle acceleration could occur via a reverse shock or from interactions between the ambient medium and jet (\citealt{matthews_2025}, \citealt{savard_2025}, \citealt{cooper_2025}). This makes deceleration a likely possibility and means the ballistic precession model is unsuitable. For a proper treatment where ambient medium interactions are important, hydrodynamic simulations are needed. These have been performed for the case of radio galaxies (\citealt{horton_2020}, \citealt{nolting_2023}) and SS~433 (\citealt{goodall_2011a}, \citealt{monceau_2014}), showing that hydrodynamic effects and deceleration play an important role in determining the morphology of the emission from precessing jets.

Secondly, as noted throughout the literature, Cir X-1 undergoes dramatic secular changes in its behaviour on timescales of decades \citep{armstrong_2013}. Cowie et al. (in prep) show that these changes likely reflect a long-term evolution in jet power, suggesting that the jet properties may not remain constant over long timescales. Furthermore, jet interaction features such as the bubbles \citep{gasealahwe_2025}, suggest that the jet was not always precessing in the past. It may be the case that Cir X-1 switches jet modes between launching a precessing jet and a fixed axis jet, or that two jets are active in the system simultaneously (Fender \& Motta 2025, in press). The observation of an URF from Cir X-1 \citep{fender_2004}, is further evidence that multiple jets with different properties exist in this source, as it is not possible to reconcile the mildly relativistic jets we observe with this URF. If multiple jet modes do exist then it becomes difficult to tell which past measurements of the jet launch axis are suitable to include when fitting a precession model. Even if multiple jet launch modes do not exist, a jet-redirecting powerful wind, launched by a precessing disc only at certain times, can result in a jet which sometimes exhibits precession and other times does not. Finally, we do not discount the possibility that that there is no steady precession with a fixed period or axis. Cir X-1, the youngest known XRB, underwent a recent cataclysmic disruption during a supernova \citep{heinz_2013}. The jet launch axis may be varying chaotically, or through other non-precession processes, as the system has not had time to settle into a steady state. Despite these facts, it seems that precession of some form, whether stable or unstable, remains the only viable option to explain the changes in the jet launch direction observed over the past 2 decades.



Despite the absence of a confident fit to a steady precession model, this could still describe the behaviour of Cir X-1. The S-shaped emission, and core emission position angle changes since 2018, can be explained by a scenario where the jet launch axis position angle has changed by an average of $\sim5\degree$ per year. Similar rates of evolution are seen in Figure~4 of \cite{tudose_2008}, from 1998 to 2003. Given that we have observed the jet axis position angle to vary by at least $110\degree$, assuming this variation arises from precession. We can then constrain the precession period to be $>10$ years, substantiated by the range of ballistic precession models fitted to the S-shape, with none having periods $<10$ years. We note an apparently faster evolution in the position angle between 2010 \citep{miller-jones_2011b} and 2011 \citep{coriat_2019}. Finally, given the S-shape and numerous observations over the course of a year (\citealt{tudose_2008}, this work) it is very unlikely we are undersampling in time, leading to an overestimate of the precession period.


The half-opening angle of the precession cone, $\psi$, can also be constrained by considering the lack of asymmetry in the VLBI images, leading to the conclusion that at this epoch, for a jet speed of $0.5c$, the instantaneous jet inclination to the line of sight must have been $i'>75\degree$, to ensure $\left(\frac{1+\beta\cos i'}{1-\beta\cos i'}\right)^3 < 2$. Therefore, throughout the precession, the instantaneous inclination to the line of sight of the jet axis must reach at least this value, meaning $\psi+i>75\degree$, where $i$ is the inclination of the precession cone to the line of sight. Furthermore, given the observed range of jet position angles, we can use the equations for the steady precession model in \cite{hjellming_1986} to form the inequality:

\begin{equation}
    90\degree-\arctan{\frac{\sqrt{\cos{2\psi} - \cos{2i}}}{\sqrt{2}\sin{i}}} > \frac{110\degree} {2} .
\end{equation}

\noindent Solving these inequalities together gives $\psi>33\degree$. If steady precession is the correct model for the jets in this system, then these constraints place Cir X-1 in a unique parameter space with respect to the few other X-ray binaries which show precession. The most relevant comparison for Cir X-1 is the system SS~433. SS~433 and Cir X-1 are both X-ray bright, peculiar X-ray binaries, embedded in their natal SNR (\citealt{bowler_2018}, \citealt{heinz_2013}). Signs of jet interactions with the ambient medium can be seen in both systems (\citealt{sell_2010}, \citealt{goodall_2011a}, \citealt{goodall_2011b}, \citealt{gasealahwe_2025}, Cowie et al., in prep). SS~433 is an older system compared to Cir X-1, aged at $10^4-10^5$ years \citep{goodall_2011a} potentially making Cir X-1 the younger analogue of SS~433. Finally, both launch relativistic jets with axes which change with time, leading to curved radio morphologies \citep{blundell_2004}. The precession period of Cir X-1 is inferred to be at least 25 times greater than the 162 day period in SS~433. Additionally, the precession half opening angle of SS~433 is  $20\degree$, at least 1.5 smaller than in the Cir X-1 case.

The constraint on the precession period rules out certain driving mechanisms. If precession is occurring we should expect some misalignment of angular momentum in the system. Evidence for a misalignment between the neutron star rotation axis and the orbital plane has recently been found through X-ray polarisation observations \citep{rankin_2024}. Misalignments in Cir X-1 are plausible, given the system was recently disrupted by a supernova explosion.

Precession of the accretion disc can drive precession of the jet, either through the launch axis of the jet being linked to the accretion disc, or through re-direction of a fixed axis jet by a powerful wind-like outflow. The accretion disc can precess due to a variety of effects \citep{caproni_2006}. Magnetically driven precession occurs when the neutron star spin is misaligned with the accretion disc angular momentum \citep{lai_1999}, however over a wide parameter space this cannot produce long period precession in neutron star systems \citep{veresvarska_2024}. The accretion disc can also precess due to tidal effects from the companion in the binary, if the accretion disc is initially misaligned with the orbital angular momentum (see \citealt{katz_1973}, \citealt{bardeen_1975}, \citealt{schandl_1994}, \citealt{fragile_2001} for various mechanisms). Given the unknowns in the system we find that tidal precession can create a period on the order of 10 years, but only for a low mass companion, large half opening angle, or a small accretion disc size (using equation~4 from \citealt{coriat_2019}). The Lense-Thirring effect can also cause precession of the accretion disc \citep{lense_1918}. However, this mechanism can only produce precession periods of a few days or less \citep{massi_2010}. The jet direction may also be set by the neutron star spin, which may also precess. This can occur either through geodetic precession due to due to misalignment with the orbital angular momentum \citep{boerner_1975} or free precession (\citealt{shaham_1977}, \citealt{fabian_1980}). Geodetic precession leads to a precession period several orders of magnitude too large to explain the jet precession and free precession is poorly understood so is difficult to assess in terms of viability.

We can compare the jets of Cir X-1 to precessing jets seen on much larger scales in radio galaxies. In this case jet precession is thought to occur through precession of the accretion disc or geodetic precession of black hole spin \citep{begelman_1980}. While the scales are vastly different, the morphology of the jets in Cir X-1 bears a striking resemblance to some radio galaxies, e.g. PKS 2300-18 \citep{misra_2025}. Furthermore, a feature common to several S-shaped radio galaxies is the existence of diffuse, lower surface brightness wings of emission, misaligned from the jet axis. We note that a qualitatively similar structure has been observed in SS~433, referred to as the radio ruff (\citealt{paragi_1999}, \citealt{blundell_2001}). However, this is interpreted to be a wind-like outflow, and is observed on $10^{-3}$ pc scales rather than the parsec scales of the diffuse emission in Cir X-1. Overall, similarities in the morphology between Cir X-1 and radio galaxies suggest a certain degree of scale invariance in these objects. However, this is contrary to theoretical expectations from considerations of the different environments these jets propagate into \citep{heinz_2002}.

Finally, the fact that jets of  Cir X-1 are still propagating mildly relativistically around 1~pc away from the system has significant implications. Over the past 2 decades, several black hole X-ray binaries have been observed to launch so-called large scale jets (see e.g. \citealt{caroteunto_2024}). However, no detection of moving large scale jets from any confirmed NSXB has been seen until Cir X-1. Other NSXBs binaries have shown resolved jets on smaller scales, e.g. the jets in Cyg X-2 observed out to $\sim60$ AU \citep{spencer_2013}. In the case of Cir X-1 the jets are still propagating and detected out to $\sim1$~pc, which is roughly the same distance as the furthest known black hole XRB jets \citep{russell_2019}. This indicates that the jets from Cir X-1 are comparable, in terms of energy, to the large scale jets from some of the most powerful black hole X-ray binaries, if the sources exist in similar ISM environments.

\section{Conclusions}

By combining the most detailed radio observations of the NSXB Cir X-1 to date, with observations in the literature going back over 20 years, we have revealed several unique features of the relativistic jets. We have found a parsec-scale S-shaped morphology in the radio emission around Cir X-1 which can be robustly attributed to relativistic jets interacting with the ambient medium. This makes Cir X-1 the second ever X-ray binary, and first confirmed neutron star X-ray binary, to show this type of jet morphology.

We have found evidence for movement of jet components within this morphology, demonstrating the jets launched by the neutron star in Cir X-1 are still propagating mildly relativistically at least $1$~pc from the system. This is the first observation of dynamic large scale jets from an accreting neutron star, and potentially implies similar jet powers to transient large scale jets launched by black hole X-ray binary systems (see also \citealt{gasealahwe_2025}).

Comparison to data going back over 20 years shows a $110\degree$ range of jet position angles on sky, the largest range of jet position angles seen from any X-ray binary, indicating the jet launch axis of Cir X-1 has changed direction by a large degree. This is the first case where this has been observed from a neutron star X-ray binary. Precession of the jets is the most likely scenario explaining this behaviour. If stable precession is occurring we can constrain the precession period to be $>10$ years and the half opening angle of the precession cone to be $>33\degree$, placing Cir X-1 in a unique area of parameter space compared to the few other X-ray binary systems with precessing jets. 

\section*{Acknowledgements}

We thank the anonymous referee for their detailed comments which helped significantly improve the manuscript. FJC was supported by STFC grant ST/Y509474/1. RF acknowledges support from UK Research and Innovation, the European Research Council and the Hintze Charitable Foundation. JvdE was supported by funding from the European Union's Horizon Europe research and innovation programme under the Marie Skłodowska-Curie grant agreement No 101148693 (MeerSHOCKS). The MeerKAT telescope is operated by the South African Radio Astronomy Observatory, which is a facility of the National Research Foundation, an agency of the Department of Science and Innovation. This work has made use of the “MPIfR S-band receiver system” designed, constructed and maintained by funding of the MPI für Radioastronomy and the Max-Planck-Society. We acknowledge the use of the ilifu cloud computing facility – www.ilifu.ac.za, a partnership between the University of Cape Town, the University of the Western Cape, Stellenbosch University, Sol Plaatje University and the Cape Peninsula University of Technology. The ilifu facility is supported by contributions from the Inter-University Institute for Data Intensive Astronomy (IDIA – a partnership between the University of Cape Town, the University of Pretoria and the University of the Western Cape), the Computational Biology division at UCT and the Data Intensive Research Initiative of South Africa (DIRISA). This work made use of the CARTA (Cube Analysis and Rendering Tool for Astronomy) software (DOI 10.5281/zen- odo.3377984 – https://cartavis.github.io). The authors would like to thank Lilia Tremou, Andrew Hughes, Francesco Carotenuto and Payaswini Saikia for scheduling the MeerKAT observations.

\section*{Data Availability}

The data availability will be subject to the ThunderKAT LSP and X-KAT XLP data
release conditions.



\bibliographystyle{mnras}
\bibliography{cirX1bib} 




\appendix

\section{Precession fit parameters}

\begin{table}
 \caption{Parameters for the ballistic precession model shown in Figure~\ref{fig:core_no_sub}. These parameters should not be interpreted as uniquely constraining.}
 \label{tab:preccession_params}
 \begin{tabular*}{\columnwidth}{l@{\hspace*{40pt}}l}
  \hline
  Command & Output \\
  \hline
  $P$, period (days) & $8160$ \\[2pt] 
  $v_j$, jet velocity & $0.7c$ \\[2pt]
  $\psi$, opening angle & $35\degree$ \\[2pt]
  $\chi$, position angle E of N & $300\degree$ \\[2pt]
  $i$, inclination & $37\degree$ \\[2pt]
  $\phi$, precession phase & $0.58$ \\[2pt]
  $s_\text{rot}$, sense of rotation & $-1$ \\[2pt]
  \hline
 \end{tabular*}
\end{table}



\bsp	
\label{lastpage}
\end{document}